\documentclass[a4paper,11pt]{article}
\usepackage{amsmath}
\usepackage{bbm}
\usepackage{cancel}
\usepackage[normalem]{ulem}
\usepackage{cite}
\usepackage[sort&compress,numbers]{natbib}

\usepackage{amsfonts}
\usepackage{amssymb}
\usepackage{latexsym}
\usepackage{xcolor}
\usepackage{url}

\newcommand{\qed}{\hfill $\Box$ \medskip}
\newcommand{\proof}{\noindent {\sc Proof:\ }}
\newcommand{\dwa}{{  2}}

\newcommand{\cztery}{{  4}}
\newcommand{\osiem}{{  8}}

\newcommand{\szesnascie}{{  16  }}

\newcommand{\pieccztery}{{  5/4}}

\newcommand{\siedemcztery}{{ 7/4}}

\newcommand{\Kerr}{\textrm{Kerr}}

\newcommand{\gKKerr}{{  g^{\Kerr,m,\vec J} K^{\Kerr,m,\vec J}}}

\newcommand{\idmJkc}{{ (\R^3, g^{\intr,m,\vec c,\vec J,k}, K^{\intr,m,\vec c,\vec J,k})}}

\newcommand{\idQk}{{ (\R^3, g^{\intr,Q,k}, K^{\intr,Q,k})}}
\newcommand{\idBQk}{{ (B(\szesnascie k), g^{\intr,Q,k}, K^{\intr,Q,k})}}

\newcommand{\gim}{{ g^{\intr, m}}}
\newcommand{\twok}{{ 2k }}
\newcommand{\fromktotwok}{{  k \le |\vec x| \le 2k }}
\newcommand{\fromonetotwo}{{ 1 \le |\vec x| \le 2 }}

\newcommand{\KvepQk}{{ K^{\intr,m,\vec J, k}}}

\newcommand{\gveQk}{{ g^{\intr,m,\vec J, k}}}

\newcommand{\rfcp}{{retarded foliation near $i^0$}}

\newcommand{\one}{{  \mathbbm{1}}}
\newcommand{\fourR}{{  2 R}}
\newcommand{\reallyfourk}{{  2 k}}

\newcommand{\notsixteen}{{  2}}
\newcommand{\hepsilon}{{  \epsilon_c}}
\newcommand{\epsilonc}{{  \epsilon_c}}

\newcommand{\nolambda}{}%{{\color{red} (nolbd)}}

\newcommand{\deltac}{{ \delta_c}}

\newcommand{\normwithnodelta }{|}

\newcommand{\intr}{\mathrm{interior}}
\newcommand{\bMinkowskimetric}{\eta}

\newcommand{\red}[1]{{\color{red}#1}}

\newcounter{mnotecount}[section]

\renewcommand{\themnotecount}{\thesection.\arabic{mnotecount}}
\newcommand{\mnote}[1]%{}
{\protect{\stepcounter{mnotecount}}$^{\mbox{\footnotesize
$%\!\!\!\!\!\!\,
\bullet$\themnotecount}}$ \marginpar{%\color{red}%
\raggedright\tiny\em
$\!\!\!\!\!\!\,\bullet$\themnotecount: #1} }

\newcommand{\ptcr}[1]{{\color{red}\mnote{{\color{red}{\bf ptc:}
#1} }}}

\newtheorem{theorem}{\sc  Theorem\rm}[section]
\newtheorem{argument}[theorem]{\sc  Evolution argument\rm}

\newtheorem{Theorem}[theorem]{\sc  Theorem\rm}

\newtheorem{Proposition}[theorem]{\sc Proposition\rm}

\newcommand{\ol}[1]{\overline{#1}{}}

\newcommand{\jlcax}[1]{}
\newcommand{\eean}{\nonumber\end{eqnarray}}

\newcommand{\og}{{\overline{g}}}

\newcommand{\fourg}{{\mathfrak g }}

\newcommand{\kk}[1]{}%{\mnote{{\bf If we consider the KK case:} #1}}

\newcommand{\beq}{\begin{equation}}

\newcommand{\FS}       %{F_1} %
                  {F}
\newcommand{\HS} %{F_2}
       {H_{\mbox{\scriptsize volume}}}

{\ptc{this should be removed in the oberwolfach version}}%

\newcommand{\ourU}{\mathbb U}%
\newcommand{\eeal}[1]{\label{#1}\end{eqnarray}}
\newcommand{\bed}{\begin{deqarr}}
\newcommand{\eed}{\end{deqarr}}
\newcommand{\bedl}[1]{\begin{deqarr}\label{#1}}
\newcommand{\eedl}[2]{\arrlabel{#1}\label{#2}\end{deqarr}}

\newcommand{\mcN}{{\mycal N}}

\newcommand{\bel}[1]{\begin{equation}\label{#1}}
\newcommand{\bea}{\begin{eqnarray}}
\newcommand{\bean}{\begin{eqnarray}\nonumber}
\newcommand{\beal}[1]{\begin{eqnarray}\label{#1}}
\newcommand{\eea}{\end{eqnarray}}

\newcommand{\nn}{\nonumber}

\newcommand{\be}{\begin{equation}}
\newcommand{\eeq}{\end{equation}}
\newcommand{\ee}{\end{equation}}
\newcommand{\beqa}{\begin{eqnarray}}
\newcommand{\eeqa}{\end{eqnarray}}
\newcommand{\beqan}{\begin{eqnarray*}}
\newcommand{\eeqan}{\end{eqnarray*}}
\newcommand{\ba}{\begin{array}}
\newcommand{\ea}{\end{array}}

\newcommand{\hyp}{\mycal S}

\newcommand{\mcM}{{\mycal M}}

\newcommand{\warn}[1]%{}%{}
{\protect{\stepcounter{mnotecount}}$^{\mbox{\footnotesize
$%\!\!\!\!\!\!\,
\bullet$\themnotecount}}$ \marginpar{%\color{red}%
\raggedright\tiny\em
$\!\!\!\!\!\!\,\bullet$\themnotecount: {\bf Warning:} #1} }

\newcommand{\R}{\mathbb R}

\newcommand{\N}{\mathbb N}

\newcommand{\eq}[1]{(\ref{#1})}

\newcommand{\ptc}[1]{\mnote{{\bf ptc:}#1}}

\newcommand{\beqar}{\begin{deqarr}}
\newcommand{\eeqar}{\end{deqarr}}

\newcommand{\beaa}{\begin{eqnarray*}}
\newcommand{\eeaa}{\end{eqnarray*}}

\newcommand{\tr}{\mbox{tr}}

\newcommand{\calM}{{\mcM}}

\renewcommand{\fourg}{{}^4g}

\DeclareFontFamily{OT1}{rsfs}{}
\DeclareFontShape{OT1}{rsfs}{m}{n}{ <-7> rsfs5 <7-10> rsfs7 <10-> rsfs10}{}
\DeclareMathAlphabet{\mycal}{OT1}{rsfs}{m}{n}

{\catcode `\@=11 \global\let\AddToReset=\@addtoreset}
\AddToReset{equation}{section}

{\catcode `\@=11 \global\let\AddToReset=\@addtoreset}
\AddToReset{figure}{section}

{\catcode `\@=11 \global\let\AddToReset=\@addtoreset}
\AddToReset{table}{section}

\renewcommand{\red}[1]{#1}
\renewcommand{\ptcr}[1]{}

\begin{document}
\title{Long time existence from interior gluing\protect\thanks{Preprint UWThPh-2016-25}}
\author{Piotr T. Chru\'{s}ciel\thanks{ University of Vienna and Erwin Schr\"odinger Institute.
{\sc Email} \protect\url{piotr.chrusciel@univie.ac.at}, {\sc URL} \protect\url{homepage.univie.ac.at/piotr.chrusciel}}}
\maketitle

\begin{abstract}
\red{
We prove  completeness-to-the-future of null hypersurfaces emanating outwards from large spheres, in vacuum space-times evolving from general asymptotically flat data with well-defined energy-momentum.
}
 The proof uses scaling and a gluing construction to reduce the problem to Bieri's stability theorem.
\end{abstract}

 \tableofcontents

\section{Introduction}
 \label{s22X16.1}

The question arises whether asymptotically flat, say vacuum,   initial data sets
lead to space-times $(\mcM,g)$ where the radiation fields can be defined.
For this one needs to be able to recede to infinity in $\mcM$ in null directions.  The object of this paper is to prove that this is indeed the case, for a large class of asymptotically \red{flat vacuum} initial data sets, under natural decay conditions on the metric and without smallness conditions.

To make things precise, consider a space-time $(\mcM,{}^4g)$ evolving out of initial data $(\hyp,g,K)$, satisfying the vacuum constraint equations, which are asymptotically flat at large distances in the asymptotic regions. We wish to address the question of existence in $\mcM$ of a family of hypersurfaces  which behave as the retarded time coordinate $u=t-r$ in Minkowski space-time. This will be modeled by a foliation by null hypersurfaces $\mcN(u)$ parameterised by a parameter $u\in (-\infty,u_0]$, so that the hypersurfaces $\mcN(u)$ intersect the asymptotically flat region of $\hyp$ in spheres which, to leading order, are coordinate spheres in manifestly asymptotically Euclidean coordinates on $\hyp$, and which recede to infinity on $\hyp$ as $u$ tends to minus infinity. We will also require that the family of spheres so obtained foliates the asymptotic region of $\hyp$, with the future-directed tangents to the generators of $\mcN(u)$ pointing outwards on $\hyp$. 
Finally, we will require that all generators of each $\mcN(u)$ are complete to the future. Such a family of null hypersurfaces will be referred to as \emph{a \rfcp}.

One can view such null hypersurfaces as being obtained from an initial asymptotically flat Cauchy surface by an infinite boost.
The  existence of a retarded foliation near $i^0$  can then be thought of as \emph{an infinite boost theorem}. We will prove such a theorem below.  
However, to avoid confusion with the already existing name associated with the Aichelburg-Sexl metrics~\cite{AS}, we will not use the \emph{infinite boost} terminology in this context.

While the \emph{finite boost theorem} has been proved a long time ago~\cite{christodoulou:murchadha}, the existence of future-complete null hypersurfaces
has only been settled so far for weak gravitational fields, or for restricted classes of initial data, or both~\cite{LindbladRodnianski2,KlainermanNicoloBook,BieriJDG,Ch-Kl,BieriZipser1}. For instance,
within the class of space-times evolving out of
vacuum asymptotically flat initial data, such foliations exist

\begin{enumerate}
  \item \label{p28VIII16.1} (obviously, by uniqueness of solutions in domains of dependence)  for  initial data which are stationary at large distances, or
  \item \label{p28VIII16.2}
  for small initial data with \emph{optimal} asymptotic conditions~\cite{BieriJDG} or,
%  \item \label{p28VIII16.3}
%  for small initial data with initial data which are Schwarzschildean to leading order~\cite{LindbladRodnianski2}.
  \item \label{p28VIII16.4}
  without smallness restrictions,
  for CMC initial data which are Schwarzschildean to high order~\cite{KlainermanNicoloBook};
  \item \label{p28VIII16.5}
  without smallness restrictions,
  for  initial data which have  well-defined total energy, momentum, angular momentum and center of mass~\cite{ChBieri}.
\end{enumerate}

So the key point of the current work is to remove  the condition of well-defined angular momentum and center of mass from the hypotheses in~\cite{ChBieri}. \red{ This is the contents of Theorem~\ref{T9V16.1} below.}
 \ptcr{the red text here and until the end of the introduction: expanded comments on the strategy of the proof}

The strategy of the proof follows closely that in~\cite{ChBieri}, but we face several new technical difficulties.  There is a standard way of reducing the proof of existence of the retarded foliation $\mcN(u)$  to a small-data existence result by scaling down. \red{This is presented in detail in the ``Evolution Argument~\ref{EA27VIII16.1}'', p.\ \pageref{EA27VIII16.1} below.}
 One can then imagine adapting the existing global existence arguments to prove directly global existence for small data in domains of dependence, but this does not appear to be straightforward within the scheme of proof of~\cite{BieriJDG}, which is the only one so far under optimal \red{decay conditions. Here optimality is with respect to the requirement of existence of a well-defined total mass and momentum of the initial data.} As in~\cite{ChBieri}, we show instead that scaled-down exterior regions can be filled-in by  initial data with small energy while preserving the vacuum constraint equations. This is done by a gluing argument which is relatively standard for initial data with well defined energy-momentum, center of mass, and angular momentum, but turns out to be rather delicate in the general case.

\red{
Now, the gluing construction requires two sets of initial data which are nearby and which will be glued together.
In our setting the exterior data are given, and the main new contribution of this work is the construction of the second, ``interior'' data set, which needs to have small energy and be near to the exterior data set in a suitable norm on the overlapping region.
It turns out that we can construct such a data set under optimal decay conditions for time-symmetric initial data, where $K_{ij}=0$. This is the heart of the proof of Theorem~\ref{T9V16.1} in the time-symmetric case, and is the contents of Proposition~\ref{PEA27VIII16.1} below.
The corresponding construction for initial data with $K_{ij}\ne 0$ is carried-out in Proposition~\ref{P27VIII16.2}, where we have not been able to handle the borderline fall-off case, and have been forced to assume   ``an $\epsilon$-amount'' of anti-parity in the initial data.
}

We emphasise that we do not make any new claims concerning regularity of the resulting ``piece of Scri''. However, the resulting space-times have enough regularity to define the radiation field, the Trautman-Bondi mass of the hypersurfaces $\mcN(u)$, and to show that the Trautman-Bondi mass tends to the ADM mass when $u$ tends to minus infinity. The reader is referred \red{to~\cite{ChBieri}}
 \ptcr{the reference "in preparation" removed, publication details for \cite{ChBieri} added}
for a discussion of these points.

\section{``Poincar\'e charges''}
 \label{s6V16.21}

\red{
A key role in the argument is played by ``Poincar\'e charges'' calculated over large spheres, and the question of the speed of their convergence as the radii of the spheres recede to infinity. It is therefore convenient to review the conditions needed to define the total  energy-momentum, angular momentum and the center of mass of  asymptotically Euclidean initial data sets. This is the aim of this section.
}

Let $\alpha\in\R^+$, $\ell \in \N$, $\ell\ge 1$.
We shall say that $(g,K)$ on the exterior $E$ of a ball in
$\mathbb{R}^3$ constitutes  an $C^{-\alpha}_\ell$-\emph{asymptotically Euclidean end}
 provided there are coordinates in which, for all
multi-indices $|\gamma|\leq \ell$, $|\beta|\leq \ell-1$,
\begin{equation} \label{af}
|\partial^{\gamma}(g_{ij}-\delta_{ij})(\vec   x)|=O(|\vec   x|^{-|\gamma|-\alpha})
 , \qquad
 |\partial^{\beta}K_{ij}(\vec   x)|=O(|\vec   x|^{-|\beta|-1-\alpha}),
\end{equation}
where $\partial$ denotes the partial derivative operator.
\red{
Unless explicitly indicated to the contrary, or otherwise clear from the context, norms such as $|\vec x|$, or $|\vec c(R)|$ in \eq{28V16.12} below, etc., are calculated using the Euclidean metric.
}
Note that the index $\ell$ refers to the
differentiability class of the metric, with $K$ being a priori only $(\ell-1)$-times differentiable.
Throughout the rest of this work we require  $\ell
\geq 4$.
We say that $(M,g,K)$ is $C^{-\alpha}_\ell$-\emph{asymptotically
Euclidean} (AE) if $M$ is the union of a compact set and a finite
number of ends, all of  which are $C^{-\alpha}_\ell$-asymptotically Euclidean.

An obvious analogue of the above are the definitions of $W^{-\alpha}_{\ell,q}$-asymptotically Euclidean manifolds and ends, where one requires
that in each end $E$ we have
\bel{20V16.4}
 g-\delta \in W^{-\alpha}_{\ell,q}(E)
 \ \mbox{ and } \ K \in  W^{-\alpha-1}_{\ell-1,q} (E)
 \,.
\ee
\red{
Here a tensor field $u$ belongs to $W^{-\alpha}_{\ell,q}(E)$  if the following norm-to-power-$q$
$$
\|u\|_{W^{-\alpha }_{\ell,q}} ^q:= \sum_{|\gamma|\le \ell} \int_{E}
 |(1+r )^{  \alpha + |\gamma| }\partial^\gamma u|^q
   \frac{d^3 x}{(1+r )^{ 3}}
% \,,
$$
is finite. We set $H^{-\alpha}_\ell := {W^{-\alpha }_{\ell,2}}$.
}

As pointed out
in~\cite{ChErice},
every $C^{-\alpha}_\ell$-asymptotically Euclidean end  with $\alpha>1/2$ possesses a
well-defined finite energy-momentum vector $(p_0,\vec p)$ when the dominant energy condition $|\vec J| \le \rho$ holds with $\rho \in L^1$. \red{The arguments there easily adapt to show that this remains true} for $W^{-\alpha}_{\ell,q}$-asymptotically Euclidean ends with $\alpha \ge 1/2$ and $q \ell>3$; compare~\cite{Bartnik86}.

Further conditions have to be imposed on the initial data to guarantee convergence of the integrals defining the centre of mass and total angular momentum.
One possible such condition is a \emph{parity requirement} (cf.~\cite[Proposition~E.1]{ChDelay}), that there exists $\alpha_- >0$ satisfying %
\bel{15VIII16.1}
 \alpha+\alpha_->2
\ee
such that  we have
\bean
&
 \big| g_{ij}(\vec   x)-g_{ij}(-\vec   x) \big|
 +(1+r)\big|\partial_k \big(g_{ij}(\vec   x)-g_{ij}(-\vec   x)\big)\big|
  =O(|\vec   x|^{ -\alpha_-})
 \,,
 &
\\
 &
 \big | K_{ij}(\vec   x)+K_{ij}(-\vec   x) \big|=O(|\vec   x|^{-1- \alpha_-})
  \,.
  &
  \label{29IV16.2+}
\eea
This requirement is in the spirit of, but weaker than the Regge-Teitelboim conditions~\cite{RT,BeigOMurchadha87} for a well-defined angular-momentum and center of mass, which are \eq{29IV16.2+} with $\alpha=1$ and $\alpha_-=2$.

The infinite-boost theorem of Section~\ref{s2V16.3} below will require some control of the integrals associated with the centre of mass and angular momentum, which we describe in detail now. Given an asymptotically flat end within an initial data set  $(\hyp, g,K)$ we set
% \ptc{sign of $J$ is the sign as that in the paper with JK multiplied by X, where
%X is the Killing vector, so maybe the four momentum has a wrong sign, but I don't care}
%
\beal{31V16.1}
  p_0(R) & = &  \frac{1}{16 \pi R} \int_{r=R}   (\partial_i g_{ij} -\partial_j g_{ii})x^j
  	\, d^2 S
   \,,
\\
 \nn
  c_k (R) & = &  \frac{1}{16 \pi R} \int_{r=R}   x^k (\partial_i g_{ij} -\partial_j g_{ii})x^j
  	\, d^2 S
\\
 && -  \frac{1}{16 \pi R} \int_{r=R}  ( g_{ki} x^i  - g_{ii} x^k) \, d^2 S
  \,,
  \label{27V16.10+}
\\
  p_k (R) & = &  \frac{1}{8\pi R} \int_{r=R} P_{ik} x^i d^2 S
   \,,
  \label{31V16.2}
\\
  J_k (R) & = &  \frac{1}{8\pi R} \int_{r=R} P_{ij} \epsilon_{k j\ell} x^\ell x^i d^2 S
  \,,
\eeal{28V16.10}
where all repeated indices are summed over. Here, and elsewhere,
$$
 P_{ij} = - K_{ij} + \tr_g K g_{ij}
 \,.
$$
The center of mass and the angular momentum are limits, as $R$ tends to infinity, of the integrals above, whenever these limits exist. While we will not require existence of the limits, we will need to assume a bound on the growth of the integrals, namely existence of $\alpha_c\in \R^+$ such that
\bel{28V16.12}
 |\vec c(R)\normwithnodelta   + |\vec J (R)\normwithnodelta   = o (R^{2-2\alpha_c})
 \,.
\ee
The constant $\alpha_c<1$
can be chosen as close to one as desired when the limits
$$
 \vec c :=\lim_{R\to\infty} \vec c(R) \ \mbox{and}\ \vec J :=\lim_{R\to\infty} \vec J(R)
$$
exist and are finite.

We claim that,
for $W^\alpha_{1,2}$-asymptotically Euclidean vacuum initial data sets with $\alpha<1$,
the largest constant $\alpha_c$ such that
\eq{28V16.12} holds satisfies
\bel{29V16.1}
 \alpha_c\ge \alpha
  \,.
\ee
For this,
consider the ``Freud superpotentials'' $\ourU^{\alpha\beta}$ defined as
\begin{eqnarray}
  \ourU^{\nu\lambda}&:= &
{\ourU^{\nu\lambda}}_{\beta}X^\beta
  + \frac 1{8\pi} \Delta^{\alpha[\nu}
  {X^{\lambda]}}_{;\alpha}
\ ,
 \phantom{xxx}
 \label{Fsup2new_new}
\\ {\ourU^{\nu\lambda}}_\beta &:= & \displaystyle{\frac{2|\det
  \bMinkowskimetric _{\mu\nu}|}{ 16\pi\sqrt{|\det g_{\rho\sigma}|}}}
g_{\beta\gamma}(e^2 g^{\gamma[\nu}g^{\lambda]\kappa})_{;\kappa}
\,,\label{Freud2.0_new}
\end{eqnarray}
where a
semicolon  denotes covariant
differentiation \emph{with respect to the Minkowski metric $\bMinkowskimetric$},
while
\bea
  \label{mas2_new}
   &
    e := \frac{\sqrt{|\det g_{\rho\sigma}|}}{\sqrt{|\det\bMinkowskimetric _{\mu\nu}|}}
     \;,
     \qquad
 \Delta^{{  \alpha}\nu}:=e\, g^{{ \alpha}\nu}- \bMinkowskimetric{}^{{ \alpha}\nu}
    \,.
   &
\eea
%%
%In vacuum we have (cf., e.g.,~\cite[Equation~(3.6)]{ChAIHP})
%%
%\bel{23X16.1}
% \partial_\alpha \ourU^{\alpha\lambda} = Q(g_{\mu\nu},\partial_\sigma g_{\mu\nu})
% \,,
%\ee
%%
%where $Q$ is a quadratic form in both space- and time-derivatives of the space-time metric  $g_{\mu\nu}$ with coefficients depending upon  $g_{\mu\nu}$, \red{as well as the vector field $X$ and its first derivatives}.
\red{Finally, $X$ is assumed to be a Killing vector field of the background Minkowski metric $\eta$.

For the purpose of the estimates  here, and  for vacuum initial data close to the Minkowskian ones, we will need an identity of the following form (cf., e.g.,~\cite[Equation~(3.6)]{ChAIHP}, compare~\cite{Changmom,ChErice})}
 \ptcr{some irrelevant text removed, refs added}
\bel{12V16.5}
 \partial_\lambda \ourU^{\nu\lambda}
  = X \big( (\partial g)^2 + K \times  \partial g + K^2\big)+  (g-\delta) \partial X (\partial g + K)
   \,.
\ee
Here, as elsewhere, $\delta$ denotes the Euclidean metric. The integrals \eq{31V16.1}-\eq{28V16.10} are essentially identical to integrals such as the left-hand side of \eq{12V16.6} below: under the asymptotic conditions used in this work, $X=\partial_\mu$ gives $p_\mu$,  $X=x^i\partial_t +t \partial_i$ gives $c_i$, and generators of rotations $x^i\partial_j - x^j\partial_i$, give $\epsilon_{ijk}J^k$.

Letting  $X$ be the generators of  rotation   or boosts, and integrating \eq{12V16.5} over a ball of radius $R$ we obtain the estimate
\bel{12V16.6}
 \int_{\{|\vec x|= R\}}  \ourU^{\nu\lambda} dS_{\nu\lambda}
  =
\int_{\{|\vec x|\le R\} }  O(|\vec x|^{-1-2\alpha})
 \le C R^{2-2\alpha}
 \,.
\ee
Integrating \eq{12V16.5} over an annulus of exterior radius $R\ge R_0 $ and interior radius $R_0$ we obtain, using weighted Sobolev embeddings,
\bean
 \int_{\{|\vec x|= R\}}  \ourU^{\nu\lambda} dS_{\nu\lambda}
  &= &
 \int_{\{|\vec x|= R_0\}}  \ourU^{\nu\lambda} dS_{\nu\lambda}
 +
\int_{\{R_0\le|\vec x| \le  R\} } o(|\vec x|^{-1-2\alpha})
\\
 &
 \le
 &
   C R_0^{2-2\alpha} + o( R^{2-2\alpha})
 \,.
\eeal{12V16.7}
Hence, as $R$ tends to infinity, for $0 < \alpha < 1$,
\bean
 |\vec c (R)\normwithnodelta   + |\vec J(R)\normwithnodelta
   & \le &
    C R_0^{2-2\alpha} + o( R^{2-2\alpha})=
  C \left(\frac{R_0}{R }\right)^{2-2\alpha} R^{2-2\alpha} + o( R^{2-2\alpha})
\\
 &
    =
  &
   o( R^{2-2\alpha})
  \,,
\eeal{12V16.8}
for all $R_0$ and $R/R_0$ sufficiently large, as desired.

An identical calculation shows that
for $W^\alpha_{1,2}$-asymptotically Euclidean vacuum initial data sets with $\alpha<1$   satisfying the parity condition \eq{29IV16.2+} we can without loss of generality assume that
\bel{29V16.2}
 \alpha_c \ge \min(\frac{1}{2} (\alpha + \alpha_-),1)
 \,,
\ee
where we have allowed $\alpha+\alpha_- <2$.

\section{The ``future-complete-null-hypersurfaces'' theorem}
 \label{s2V16.3}

In this section we prove the \emph{future-complete-hypersurfaces theorem} for a large class of AE initial data.
The assumptions in the time-symmetric case appear to be optimal with respect to the definition of total mass. In the case where $K\not\equiv 0$ our proof requires a very mild parity assumption to eliminate  the borderline case $\alpha_-=\alpha=1/2$, where
  $\alpha_-$ is the anti-parity  exponent of \eq{29IV16.2+}. The hypothesis is used to obtain slightly better control of the growth of the center-of-mass and angular-momentum integrals when $\alpha=1/2$, and we
expect it to be unnecessary:

\begin{Theorem}
  \label{T9V16.1}
  Let $(\hyp,g,K)$ be an $H^{-\alpha}_\ell$-AE  initial data set with $\alpha \ge 1/2 $, $\ell\ge 4$, and timelike four-momentum.
  \ptcr{I confirm that $H^{-\alpha}_\ell$ is correct here}
  Assume that  $(\hyp,g,K)$   is vacuum at large distances. Suppose that either
\begin{enumerate}
  \item \label{P2T9V16.1}
  the initial data are time-symmetric, i.e.\ $K\equiv 0$, or
  \item
  \label{P3T9V16.1}
  the anti-parity  exponent $\alpha_-$ in \eq{29IV16.2+} satisfies
  \bel{29V16.3}
   \alpha_- > \frac{1}{2}
    \,.
  \ee
\end{enumerate}
   Then the \red{vacuum} maximal globally hyperbolic development of $(\hyp,g,K)$ contains
   a \rfcp, as defined in Section~\ref{s22X16.1}.
\end{Theorem}

\proof
We start by noting that the proof has two components: the first is an interior gluing statement, the second a uniqueness-in-domains-of-dependence property of the evolution problem.

For the purpose of some arguments that will follow we introduce a natural number $k\in \N$ which will need to be taken large.

For all $\varepsilon$ sufficiently small, consider the initial data $(\R^3\setminus B(2), g_{\varepsilon,k}, K_{\varepsilon,k})$, obtained by scaling-down the
complement of a coordinate ball of radius $16 k/\varepsilon$ in an asymptotically Euclidean end of $(\hyp,g,K)$ (compare~\eq{12V16.10} below). We will  show  that for all $\varepsilon$ small enough and $k$ large enough the data
$(\R^3\setminus B(2), g_{\varepsilon/k}, K_{\varepsilon/k})$ can be extended, by gluing, to a vacuum data set, say $(\R^3,\hat g_{\varepsilon,k},\hat K_{\varepsilon,k})$, with small weighted Sobolev norms so that the evolution theorem of Bieri~\cite{BieriJDG} applies.

The following evolution argument will be common to  both cases 1. and 2.:

\begin{argument}\label{EA27VIII16.1} {\rm
Making  $\varepsilon $ smaller and $k$ larger if necessary, the vacuum solution, say $(\calM,\fourg_{\varepsilon,k})$ associated with the
initial data set $(\R^3,\hat g_{\varepsilon,k},\hat K_{\varepsilon,k})$ exists globally by~\cite{BieriJDG} and contains a foliation by null hypersurfaces $\mcN(u)$ defined  by a retarded null coordinate $u\in \R$. Uniqueness of solutions within domains of dependence guarantees that the space-time metric in the domain of dependence of $(\R^3\setminus B(16 k/\varepsilon),g,K)$ within the
space-time $(\mcM,\fourg)$ obtained by evolving $(\hyp,g,K)$ will, after a
constant rescaling of the space-time metric, be isometric to the domain of
dependence of $(\R^3,\hat g_{\varepsilon,k},\hat K_{\varepsilon,k})$ within  $(\calM,\fourg_{\varepsilon,k})$, and will contain the hypersurfaces $\mcN(u)$ with $u\in (-\infty, u_0]$ for some $u_0\in \R$, forming the required \rfcp.}
\qed
\end{argument}

\red{Before continuing the proof of Theorem~\ref{T9V16.1},} some further generalities are in order.
To avoid an unnecessary discussion of logarithms that could arise in some integrals, without loss of generality we can, and will, assume that
$$
 \alpha < 1
 \,.
$$

Set
$$
 E(R):= \R^3\setminus B(R)
 \,,
  \quad
 A(R) =\{R
 \le |\vec x\normwithnodelta  \le \fourR\}
 \, .
$$

For all $\varepsilon$ sufficiently small consider the family of scaled initial data sets
$(E(1), g^\varepsilon,K^\varepsilon)$ defined, in local coordinates on $E(1)$
as
\bean
 &
 g^\varepsilon_{ij} (\vec x):= g_{ij}(\vec x/\varepsilon)= \delta_{ij}
  + \varepsilon^{-\alpha}  o(|\vec x|^{-\alpha})
 \,,
  &
\\
 &
 K^\varepsilon_{ij} (\vec x):= K_{ij}(\vec x/\varepsilon)=
    \varepsilon^{-\alpha-1}  o(|\vec x|^{-\alpha-1})
 \,.
  &
\eeal{12V16.10}
Let $\epsilon>0$, for all $\varepsilon$ small enough we will have
\bel{12V16.11}
 \| g^\varepsilon-
 \delta\|_{W^{-\alpha}_{\ell,2}(E(1))}
 +
 \| K^\varepsilon\|_{W^{-\alpha-1}_{\ell,2}(E(1))}
  \le
   \frac \epsilon 3
   \,.
\ee
%%\ptc{beginning of generalWithBowenYork.tex}

We will write
$$
 \mathring Q(R_0)\equiv (\mathring p_0(R_0), \vec{ \mathring p} (R_0),
 \vec{ {\mathring c}}(R_0),\vec{ \mathring J}(R_0))
$$
for the charge integrals at radius $R_0$ associated with the unscaled metric $g$.
The following scaling properties of the global charges \eq{31V16.1}-\eq{28V16.10} are easily derived, as calculated for the initial data $(g^{\varepsilon,R},K^{\varepsilon,R}):=(g^{\varepsilon/R},K^{\varepsilon/R}) $:
\bean
  p_0(R_0)  & := &  \frac{1}{16  \pi R_0} \int_{r=R_0}   (\partial_i g^{\varepsilon,R}_{ij} -\partial_j g^{\varepsilon,R}_{ii})x^j
  	\, d^2 S
  \,,
\\
 & = & \frac{\varepsilon }{R}
  \,
  \mathring p_0\left(\frac{RR_0 }\varepsilon\right)
   \,,
 \label{pVIII16.3}
\\
 \nn
  c{}^\ell (R_0) & : = &  \frac{1}{16  \pi R_0} \int_{r=R_0}   x^\ell (\partial_i g^{\varepsilon,R}_{ij} -\partial_jg^{\varepsilon,R}_{ii})x^j
  	\, d^2 S
\\
 && -  \frac{1}{16 \pi R_0} \int_{r=R_0}  ( g^{\varepsilon,R}_{\ell i} x^i  - g^{\varepsilon,R}_{ii} x^\ell) \, d^2 S
  \nn
\\
 & = &   \left(\frac{\varepsilon }{R}\right)^2 {\mathring c}{}^\ell\left(\frac{RR_0 }\varepsilon\right)
  \,,
  \label{p27V16.10+1}
\\
  p_\ell (R_0) & := &  -\frac{1}{8\pi R_0} \int_{r=R_0} P^{\varepsilon,R}_{i\ell} x^i d^2 S
  \nn
\\
 & = &    \frac{\varepsilon }{R}
  \,
  \mathring p_\ell\left(\frac{RR_0 }\varepsilon\right)
   \,,
  \label{p31V16.2+1}
\\
  J_\ell (R_0) & := &  \frac{1}{8\pi R_0} \int_{r=R_0} P^{\varepsilon,R}_{ij} \epsilon_{\ell j h} x^i x^h d^2 S
  \nn
\\
 & = &   \left(\frac{\varepsilon }{R}\right)^2 \mathring J_\ell\left(\frac{RR_0 }\varepsilon\right)
  \,.
\eeal{p28V16.10+1}

We want to glue the  initial data \eq{12V16.10} with suitable interior initial data on an annulus $A(R)$, for $\varepsilon$ small enough and  for $R$ large. For this we need to construct a  family of ``interior data'' on $\R^3$, which will be denoted by $(\hat g, \hat K)$,  with well controlled charge integrals.
It turns out that this can be carried-out with the choice%
\bel{p13V16.2}
 R \mbox{\ a multiple of\ }  k
% \,.
\ee
for all $k$ large enough. More precisely, we will use $R=\dwa k$ in the static case, $R=\cztery k$ in the parity-symmetric case, and $R=\osiem k$ in the general case. In view of \eq{p13V16.2} and \eq{12V16.8}, to do the matching of the interior solution to the exterior one will need
\bea
&
 |\vec c\normwithnodelta   + |\vec J\normwithnodelta
 =
   o( R^{2-2\alpha_c}) =  o( k^{2-2\alpha_c})
  \,,
  &
\eeal{p12V16.3-p}
with
\bel{27VIII16.6}
   \alpha_c= \frac 12 (\alpha+\alpha_-)\ge \alpha
\ee
(compare \eq{29V16.1}).
We will therefore consider initial data $(\hat g, \hat K)$ with charge parameters $Q$ such that
\bea
&
  \frac 12 \varepsilon |\mathring m|\le |m| \le 2 \varepsilon |\mathring m|
 \,,
 \quad
  |\vec p| \le 2  \varepsilon |\vec{\mathring p}|
  \,,
  \quad
 |\vec c\normwithnodelta   + |\vec J\normwithnodelta
 \le \lambda   k^{2-2\alpha_c}
  \,,
  &
\eeal{p12V16.3p}
where $(\mathring m, \vec{\mathring p}\,)\equiv\mathring p $ is the ADM four-momentum of $(\hyp,g,K)$,
with a constant $0<\lambda\le 1$ equal to one if $\alpha_c>\frac 12$, and which will be chosen very small if $\alpha_c=\frac 12$.

We pass now to the static case which is simpler, and where a sharper result is established. As before, the first step is a gluing construction:

\begin{Proposition}
 \label{PEA27VIII16.1}
  Under the hypotheses of point~\ref{P2T9V16.1}.\ of Theorem~\ref{T9V16.1}  let moreover $1/2\le \alpha<1$. There exists  a sequence of  scalar flat metrics $g^k$ on $\R^3$ which coincide  with $g^\varepsilon$ outside of a ball of radius $\cztery k$,
with $\|g^k-g^\varepsilon\|_{W^{-\alpha}_{\ell,2}}$ tending to zero as $k$ tends to infinity, and with
\bel{27VIII16.2}
 \|g^k-\delta\|_{W^{-\alpha}_{\ell,2}(\R^3)}\to_{\varepsilon\to0}  0
 \,.
\ee
\end{Proposition}

\noindent{\sc Proof of Proposition~\ref{PEA27VIII16.1}:}
Let $(\R^3,\gim  )$ be a family of parity-symmetric scalar-flat metrics smoothly varying with $m$ in an interval $[0,m_0)$, with $m_0>0$, which for definiteness we take to coincide with the space-Schwarzschild metric outside of $B(R_0)$ (compare~\cite{ChDelay2}).  We also assume that for every $R$ the $C^\ell(B(R))$-norm of $\gim -\delta$ goes to zero as $m$ goes to zero.
Decreasing $m_0$ if necessary we can also arrange to have
\bel{p12V16.12}
 \| \gim -\delta\|_{W^{-\alpha}_{\ell,2}(\R^3 )}
  \le
   \frac \epsilon 3
   \,,
\ee
for all $m\in[0,m_0)$. Existence of such families follows e.g.\ from the proof of~\cite[Proposition~5.1]{ChBieri}.

Let $\vec a \in \R^3$ and let  $(\R^3,g^{\intr, m,\vec a} )$ be the family of metrics obtained by applying a translation by the vector $-\vec a$
to $ \gim   $ in a coordinate system in which the metric takes the usual explicit conformally Euclidean form. Then   $(\R^3,g^{\intr, m,\vec a} )$ is Ricci-scalar flat, has ADM energy $p_0=m$ and center-of-mass $\vec c =m\vec a$.

From now on we work in the region $R\ge   |\vec a| + R_0$,
where we have
\bel{p31V16.3}
g^{\intr, m,\vec a} = \Big(\underbrace{1+\frac{m}{2|\vec x - \vec a|}}_{=:\phi_{m,\vec a }} \Big)^4\delta
\,.
\ee
To simplify notation, we will write $\ol g$ for $g^{\intr, m,\vec a} $.

Assuming $|\vec a|+|m|\le R/4$, % and writing $\phi$ for $\phi_{m,\vec a }$,
the integrals \eq{31V16.1}-\eq{27V16.10+} for the metric $\ol g$ read
%\ptcheck{11VIII\notsixteen : p0 is ok, and the formal integral for cj, the last line has I believe been checked by mathematica}
%
\bea
  p_0(R)
%  & = &
%  \frac{1}{\notsixteen  \pi R} \int_{r=R}   (\partial_i g_{ij} -\partial_j g_{ii})x^j
%  	\, d^2 S
%\\
 \label{p31V16.1a}
  & = &  -\frac{1}{2 \pi R} \int_{r=R}  \phi_{m,\vec a }^3   x^j\partial_j \phi_{m,\vec a }
  	\, d^2 S
 \,,
\\
 \label{p31V16.1a+}
  & = &  m\Big(1 + O\big((m+ |\vec a|) R^{-1}\big)\Big)
%  =  m + O\big((m^2+  |\vec c| )R^{-1}\big)
 \,,
\\
 \label{p31V16.5}
  c{}^j (R) & = &  \frac{1}{8 \pi R} \int_{r=R}   x^j \phi_{m,\vec a }^3 (\phi_{m,\vec a } - 4 x^i\partial_i \phi_{m,\vec a })
  	\, d^2 S
\\
 &=&
  m a^j \big(1+     O\big((m+ m |\vec a|
 ) R^{-1}\big) \big)
  \,.
\eeal{p28V16.10a}

Let $k\in \N$ be a  number satisfying $k\ge R_0$, which will be soon taken to be very large.
Let $\chi\in C^\infty(\R)$ satisfy \red{$|\chi'|\le C$ for some constant $C$ and}
\bel{p12VIII16.2}
 \mbox{$\chi(x)=0$ for $x\le \pieccztery$, $\chi(x)=1 $ for $x\ge \siedemcztery$, and $0\le \chi \le 1$.
 }
\ee
Set
\bean
 \lefteqn{
 \chi_k (\vec x)= \chi  \left( \frac {|\vec x|}k \right)
\quad
 \Longrightarrow
 } &&
\\
 &&
% \quad
 \chi_k(\vec x) =\left\{
           \begin{array}{ll}
             0, & \hbox{$|\vec x | \le   k $;} \\
             1, & \hbox{$|\vec x | \ge \dwa k$,}
           \end{array}
         \right.
\ \mbox{and} \
 |\nabla \chi_k(\vec x)| \le \frac{C  }{k  } \one_{A(k)}(\vec x) \le \frac{2C  }{ |\vec x| }
 \,.
 \phantom{xx}
\eeal{p18VIII16.8}
Here and elsewhere we use the letter $C$ to denote a possibly large positive constant which may vary from line to line, \red{and $\one_{\Omega}$ denotes the characteristic function of a set $\Omega$}.

We restrict the range of translation vectors $\vec a$ to  vectors satisfying
\bea
&
 |\vec a \normwithnodelta  \le \lambda k^{2-2\alpha_c} = \left\{
                                 \begin{array}{ll}
                                   k^{2-2\alpha_c}, & \hbox{$\alpha_c>1/2$;} \\
        \lambda  k, & \hbox{$\alpha_c=1/2$.}
                                 \end{array}
                               \right.
  &
\eeal{p12V16.3-p-}

For $k\in\N$ we scale down all metrics from $A(\notsixteen k)$ to $A(1)$, setting for $\vec x \in A(1)$
\bean
 g^{k,\varepsilon}_{ij} (\vec x)
  &:=  &
   g_{ij}( \notsixteen  k \vec x/\varepsilon)= \delta_{ij}
  + o(k^{-\alpha}\varepsilon^{ \alpha} )
 \,,
\\
 \ol g^{k}_{ij} (\vec x)
  & := &
    \ol g_{ij}(\notsixteen  k \vec x )=  \Big(\underbrace{1+\frac{m}{2|\notsixteen k \vec x - \vec a|}}_{=:\phi_{m,\vec a,k }} \Big)^4\delta_{ij}
 \nn
\\
  & = &  \delta_{ij}
  + O(mk^{-1} + \lambda m k^{ -2\alpha_c} ) = \delta_{ij}
  + O(mk^{-1})
 \,.
\eeal{p12VIII16.1}
Let $\chi$ be as in \eq{p12VIII16.2}. Working still on $A(1)$, the metric
$$
 \chi  g^{k,\varepsilon}_{ij} + (1-\chi)  \ol g^{k}
$$
has scalar curvature which is
\bel{p14VIII16.1}
 O(mk^{-1}  )    + o(k^{-\alpha}\varepsilon^{ \alpha} )
  \,.
\ee
%.
Hence, using the results of~\cite{ChDelay,ChDelayHilbert},  for $k$ large enough
we can correct the metric by terms of order as in \eq{p14VIII16.1}
to obtain a metric, denoted by $g^k$, which is scalar-flat up to the projection on the space spanned by the functions $\{1,x^i\}_{i=1}^3$.
It then remains to show that this last projection vanishes as well under a judicious choice of $m$ and $\vec a$. This will follow  by the usual fixed-point arguments after an analysis of the ``balance formula'' (recall that $\ourU$ has been defined in \eq{Fsup2new_new}):
\begin{eqnarray}
  \frac 12 \int_{\{|\vec x|=\dwa\}}
 \ourU^{\alpha\beta}dS_{\alpha\beta}
  -  \frac 12 \int_{\{|\vec x|=1\}}
 \ourU^{\alpha\beta}dS_{\alpha\beta}
 =
  \frac 12 \int_{\{\fromonetotwo\}}
 \partial_\alpha \ourU^{\alpha 0 } d\mu_{g^k}\,,
\label{p12VIII16.3}
\end{eqnarray}
for the metrics $g^k$, where the vector field $X$ arising in the definition of $\ourU^{\alpha\beta}$ is $X=\partial_t$ or $X=t\partial_i+ x^i \partial_t$. For the sake of the estimates it is convenient to use the equivalent identities of~\cite[Section~3]{CCI}, which do not invoke space-time fields.

When $X=\partial_t$, the left-hand side of \eq{p12VIII16.3} is $p_0(\dwa)-p_0(1)$. Now, near ${\{|\vec x|=1\}}$ the metric $g^k$ coincides with $\ol g^k$, and a rescaled version of  \eq{p31V16.1a}-\eq{p31V16.1a+} with $R=\notsixteen k$ gives
\bea
  p_0(1)
%  & = &
%  \frac{1}{\notsixteen  \pi R} \int_{r=R}   (\partial_i g_{ij} -\partial_j g_{ii})x^j
%  	\, d^2 S
%\\
 \label{p12VIII16.5}
  & = &   \frac{m}{\notsixteen  k}  + O ( {m^2}{k^{-2}}+  \lambda m k^{ -2\alpha_c}   )
 \,.
\eea

Next, near ${\{|\vec x|=\dwa \}}$ the metric $g^k$ coincides with $  g^{k,\varepsilon}$. Recall that $\mathring m$ denotes the ADM mass of $g$. A calculation similar to that leading to \eq{12V16.8} together with scaling shows that
\bea
  p_0(\dwa)
%  & = &
%  \frac{1}{\notsixteen  \pi R} \int_{r=R}   (\partial_i g_{ij} -\partial_j g_{ii})x^j
%  	\, d^2 S
%\\
 \label{p12VIII16.7}
  & = &   \frac{\varepsilon \mathring m}{\notsixteen  k} + o(\varepsilon^{2\alpha}k ^{-2\alpha })
 \,.
\eea

The right-hand side of \eq{p12VIII16.3} is the sum of the $L^2$-projection operator  on the function $1$  of the scalar curvature $R(g^k)$ of the metric $g^k$, and of error terms which are quadratic in the first derivatives of the metric:
$$
 \int_{\{\fromonetotwo \}}
  \left( R(g^k)
 +
   O(m^2k^{-2} )+o(k^{-2\alpha} \varepsilon^{2\alpha})
 \right)
 d\mu_{g^k}\,.
%\label{p12VIII16.8}
$$
From \eq{p12VIII16.3}, taking into account the boundary terms (compare \eq{p31V16.1a+}), one finds
\bean
 \lefteqn{
 \notsixteen  k \int_{\{\fromonetotwo \}}
    R(g^k)
 d\mu_{g^k}
  =
 {\varepsilon \mathring m - m }
 }
 &&
\\
 &&
 + O(m \lambda k^{1-2\alpha_c}
 +
 m^2k^{-1}
% +m^2\lambda^2 k^{ 1-4\alpha_c}
 )
 +o(k^{1-2\alpha} \varepsilon^{2\alpha})
   \,.
\label{p12VIII16.10}
\eea

When $X=x^i\partial_t+ t \partial_i$, the left-hand side of \eq{p12VIII16.3} is $c{}^i(\dwa)-c{}^i(1)$. As already pointed out, near ${\{|\vec x|=1\}}$ the metric $g^k$ coincides with $\ol g^k$. To calculate $c{}^i(1)$, recall that
\bea
  c{}^j (1) & = &  \frac{1}{8 \pi  } \int_{|\vec x|=1}   x^j (\phi_{m/R,\vec a/R}^4 - 4 \phi_{m/R,\vec a/R }x^i\partial_i \phi_{m/R,\vec a/R })
  	\, d^2 S
  \,.
 \phantom{xxx}
\eeal{p28V16.10a+}
By parity considerations, the integral will remain unchanged if $\phi_{m/R,\vec a/R}^4$ in the integrand is replaced by
$$
 \frac 12 \left( \phi_{m/R,\vec a/R}^4 - \phi_{m/R,-\vec a/R}^4 \right)
  \,.
$$
A calculation shows that on $A(1)$ we have, for $(m+|\vec a|)/R < 1/2$,
\bel{p15VIII16.11}
  \frac 12 \left( \phi_{m/R,\vec a/R}  - \phi_{m/R,-\vec a/R}  \right) =     m  \frac{ \vec a \cdot \vec x} {4 R^2 |\vec x |^3} + O(m |\vec  a|^2 R^{-3})
  \,,
\ee
leading to
\bel{p15VIII16.13}
  \frac 12 \left( \phi_{m/R,\vec a/R}^4 - \phi_{m/R,-\vec a/R}^4 \right) =    m  \frac{ \vec a \cdot \vec x}{R^2 |\vec x |^3}
   + O(m^2 |\vec  a|  R^{-3} + m |\vec  a|^2 R^{-3})
  \,.
\ee
Similarly,  the integral will remain unchanged if $\phi_{m/R,\vec a/R}^3x^i\partial_i \phi_{m/R,\vec a/R } $ in the integrand is replaced by
$$
 (*):=\frac 12 \left( \phi_{m/R,\vec a/R}^3 x^i\partial_i \phi_{m/R,\vec a/R } - \phi_{m/R,-\vec a/R}^3 x^i\partial_i \phi_{m/R,-\vec a/R } \right)
  \,.
$$
The estimates of the error terms arising in $(*)$, needed in the calculations, are best carried-out using  formulae such as
$$
 (*) =\frac 12\int_{-1}^1 \frac{d
 \left( \phi_{m/R,s\vec a/R}^3 x^i\partial_i \phi_{m/R,s\vec a/R } \right)
 }{ds} ds
  \,.
$$
%,
After some work one finds, still on $A(1)$ and with $(m+|\vec a|)/R < 1/2$,
\bel{p15VIII16.12}
 (*) =   5 m \frac{  \vec a \cdot \vec x}{R^2 |\vec x |^3} + O(m |\vec  a|^2 R^{-3}+m^2 |\vec  a|  R^{-3})
  \,,
\ee
so that
\bea
  c{}^i(1)
%  & = &
%  \frac{1}{\notsixteen  \pi R} \int_{r=R}   (\partial_i g_{ij} -\partial_j g_{ii})x^j
%  	\, d^2 S
%\\
 \label{p15VIII16.13+}
  & = &    \frac{m a^i}{ R^2}
   + O(m^2 |\vec  a|  R^{-3} + m |\vec  a|^2 R^{-3})
 \,.
\eea

Recall that ${\mathring c}{}^i(R)$ denotes the center-of-mass integral at radius $R$ of $g$.
As near ${\{|\vec x|=\dwa \}}$ the metric $g^k$ coincides with $  g^{k,\varepsilon}$, the center of mass at radius $\dwa$ of the metric $g^k$ is
\bea
  { c}{}^i(\dwa)= \varepsilon^2  R^{-2}{\mathring c}{}^i(\dwa R\varepsilon^{-1})
 \,.
\eeal{p18VIII16.2}
The right-hand side of \eq{p12VIII16.3} is the sum of the $L^2$-projection operator  on the function $x^i$  of the scalar curvature $R(g^k)$ of the metric $g^k$, and of error terms which are quadratic in the first derivatives of the metric:
$$
 \int_{\{\fromonetotwo \}}
  \left( x^i  R(g^k)
 +
   O(m^2k^{-2}
 %+\lambda^2 m^2 k^{ -4\alpha_c}
)
+o(k^{-2\alpha_c} \varepsilon^{2\alpha})
 \right)
 d\mu_{g^k} \,,
%\label{p12VIII16.8}
$$
where for $\alpha_c>\alpha$ we have used the anti-parity condition on the metric to obtain that the integral of $x^i o(k^{-2\alpha} \varepsilon^{2\alpha})$ produces an error which is $o(k^{-2\alpha_c} \varepsilon^{2\alpha})$.
From \eq{p12VIII16.3} we conclude, keeping in mind that the scaling parameter $R$ (not to be confused with the scalar curvature $R(g^k)$) equals $R=\notsixteen  k$,
\bean
 \lefteqn{
 (\notsixteen k)^{ 2  } \int_{\{\fromonetotwo \}}
    x^i  R(g^k)
 d\mu_{g^k}
  =
 {\varepsilon^2   {\mathring c}{}^i(\cztery k\varepsilon^{-1}) -  m a^i }
 +o(k^{2-2\alpha_c } \varepsilon^{2\alpha})
%\phantom{xxxx}
 }
 &&
\\
 &&
\phantom{xxxx}
 + O(
 m^2
 +m^2\lambda^2 k^{2-4\alpha_c}
 + m^2 \lambda k^{1-2 \alpha_c }
 + m \lambda^2 k^{3-4 \alpha_c }
 )
   \,.
\label{p15VIII16.21}
\eea
We rescale $m\in (\varepsilon \mathring m/2, 2 \varepsilon \mathring m)$ to a mass parameter $m_s$ belonging to the interval $(\mathring m/2,2 \mathring m)$:
\bel{20IX16.2}
 m = m_s \varepsilon
 \,.
\ee
We rescale $\vec c=m \vec a$ to a vector $\vec c_s$ belonging to a unit ball:
\bel{pVIII16.1}
 \vec c_s  := \frac{m \vec  a }{\lambda R^{2-2\alpha_c}}
  \,,
   \quad
  \vec{\mathring{c}}_s (R)  := \frac{\vec{\mathring  c}(\dwa R\varepsilon^{-1})}{\lambda (\dwa R)^{2-2\alpha_c}}
 \,,
\ee
thus
$$
 \mbox{$|\vec c_s|\le 1$ and $| \vec{\mathring{c}}_s  (R)|\to_{R\to\infty}0$.
  }
$$
In terms of these variables, \eq{p12VIII16.10} and \eq{p15VIII16.21} can be rewritten as
\bean
 \lefteqn{
 \frac{\notsixteen  k}{\varepsilon} \int_{\{\fromonetotwo \}}
    R(g^k)
 d\mu_{g^k}
  =
 {\mathring m - m_s }
 }
 &&
\\
 &&
 + O( \lambda k^{1-2\alpha_c}
 +
 \varepsilon k^{-1}
% +m^2\lambda^2 k^{ 1-4\alpha_c}
 )
 +o(k^{1-2\alpha} \varepsilon^{2\alpha-1})
   \,,
\label{20IX16.1}
%\\
\eea
\bean
 \nn
\lefteqn{
\lambda^{-1} (\notsixteen k)^{ 2\alpha_c  } \int_{\{\fromonetotwo \}}
    x^i  R(g^k)
 d\mu_{g^k}
  =
 {\varepsilon^2   {\mathring c}_s^i(\dwa k) -  c_s^i }
 +o(\lambda^{-1} \varepsilon^{2\alpha})
 }
 &&
\\
 &&
 + O( \varepsilon^2
% m^2
(\lambda^{-1} k^{ 2 \alpha_c -2}
 +\lambda  k^{-2\alpha_c}
 + k^{-1})
 + \varepsilon %m
    \lambda  k^{1-2 \alpha_c }
 )
   \,.
 \label{pVIII16.5}
\eea
We are now ready to show that for $ k$ large enough we can choose $(m,\vec c_s)$ so that the right-hand sides of \eq{p12VIII16.10} and \eq{pVIII16.5} vanish:
Consider the sequence of maps, denoted by $\Phi_k(m,\vec{  c}_s)$, which to
$$
 (m_s,\vec{  c}_s)\in
 (\frac 12  \mathring m,
 2  \mathring m)
 \times B(1)\subset \R\times \R^3
$$
assign
$$
 - \Big(
 \frac{\notsixteen  k} \varepsilon \int_{\{\fromonetotwo \}}
    R(g^k)
 d\mu_{g^k}
- \mathring m
 \,,
\lambda^{-1} (\notsixteen k)^{ 2\alpha_c  } \int_{\{\fromonetotwo \}}
    x^i  R(g^k)
 d\mu_{g^k})
 \Big) \in \R\times \R^3
 \,.
$$
Choose $\lambda =\varepsilon^{\alpha}$.
Let $\epsilon_1>0$, we can choose  $\varepsilon$ small enough
so   that the integrals  of the error term  $O(\lambda  k^{1-2 \alpha_c }
 )$  in \eq{20IX16.1}-\eq{pVIII16.5} are each smaller than $\epsilon_1/2 $. We can then choose $k$ large enough so that the sum of all remaining error terms in \eq{20IX16.1}-\eq{pVIII16.5} is smaller than $\epsilon_1/2 $. But then the maps $\Phi_k$ differ  from the identity by a multiple of $\epsilon_1$ in the sup norm. From e.g.\ Lemma 5.2 in~\cite{HSW} one concludes that the images of all the $\Phi_k$'s, with $k$ sufficiently large and $\varepsilon$ (and hence $\epsilon_1$) sufficiently small, contain the origin in $\R\times \R^3$. Equivalently, there exists a choice of  parameters $(m,\vec a)$ so that $g^k$ is Ricci-scalar flat for all $k$ large enough.

By construction, it follows from \eq{12V16.11} and \eq{p12V16.12} that \eq{27VIII16.2} holds, which finishes the proof of Proposition~\ref{PEA27VIII16.1}.
\qed

\red{We are ready now to prove point~\ref{P2T9V16.1}.\ of Theorem~\ref{T9V16.1}.} It follows from Proposition~\ref{PEA27VIII16.1} that the norm of the data $(\R^3,g^k)$, as needed for the stability theorem in~\cite{BieriJDG}, can be made smaller than $\epsilon$ by decreasing $\varepsilon$ and increasing $k$ if necessary. The initial data $(\hat g_{\varepsilon,k},\hat K_{\varepsilon,k}) $ invoked in the Evolution Argument~\ref{EA27VIII16.1}, p.~\pageref{EA27VIII16.1} are defined to be  $(\hat g_{\varepsilon,k},\hat K_{\varepsilon,k}):=(g^k,0)$.  Point~\ref{P2T9V16.1}.\ of Theorem~\ref{T9V16.1} follows now from that argument.

As should be clear by now,  point~\ref{P3T9V16.1}.\  follows similarly from the Evolution Argument~\ref{EA27VIII16.1} together with the following gluing result, which will complete the proof of Theorem~\ref{T9V16.1}:
\qed

\begin{Proposition}
 \label{P27VIII16.2}
  Under the hypotheses of point~\ref{P3T9V16.1}.\ of Theorem~\ref{T9V16.1}, assume moreover that $\alpha<1$. There exists  a sequence of vacuum initial data $(\R^3,g^k,K^k)$  which coincide  with $(g^\varepsilon,K^\varepsilon)$ outside of a ball of radius $\szesnascie k$ and satisfy
\bel{27VIII16.4}
 \|(g^k-g^\varepsilon,K^k-K^\varepsilon)
  \|_{W^{-\alpha}_{\ell,2}\oplus W^{-\alpha-1}_{\ell-1,2}}\to_{k\to\infty} 0
   \,,
\ee
as well as
\bel{27VIII16.3}
 \|(g^k-\delta,K^k )
    \|_{W^{-\alpha}_{\ell,2}\oplus
        W^{-\alpha-1}_{\ell-1,2}}\to_{\varepsilon\to0}  0
 \,.
\ee
\end{Proposition}

\noindent{\sc Proof of Proposition~\ref{P27VIII16.2}:}
Our first step will be to construct a family of small-energy interior initial data sets with zero center of mass and zero momentum, but with the required range of masses and angular momenta. For this, let $\vec n$ be a  unit vectors in Euclidean $\R^3$ and let $\sigma_1(\vec n,\vec x)$  and $\tau_1(\vec n, \vec x)$ denote the solutions of the linearised vacuum constraint equations, supported in $A(1)$,
as described
in~\cite{HSW} (see Proposition~3.1 there),
 which are used to construct initial data sets with  angular momentum $ \vec n$ modulo an error as small as desired, as made precise in that reference.
Note that the tensors $\sigma_1(\vec n, \cdot)$  are obtained by simply rotating the coordinate system when  rotating $\vec n$, and therefore their Sobolev norms of any order are independent of $\vec n$; similarly for  $\tau_1(\vec n, \cdot)$.

Recall that $\alpha_c$ has been defined in \eq{27VIII16.6}.
Decreasing $\alpha_-$ if necessary we can, and will, assume that $1/2<\alpha_-<1$, and thus also $1/2<\alpha_c<1$.

For $\vec J \ne 0$  set
\bea
&
 \sigma_{\vec J}^k(\vec x)
  := % k^{\lambdac- \frac 12}
   |\vec J|^{\frac 12}\sigma_1\Big(\frac{\vec J}{|\vec J|},\frac{\vec x }{k}\Big)
  \,,
  \quad
% \hat \sigma_{\vec c}^k(\vec x)
%  = |\vec c\normwithnodelta \, \hat \sigma_1\Big(\frac{\vec c}{|\vec c|},\frac{\vec x }{k}\Big)
%  \,,
%&
%\\
% &
  \tau^k_{\vec J}(\vec x)
   :=  |\vec J \normwithnodelta ^{\frac 12}{k^{\frac 12 -\lambda_c}}\, \tau_1\Big(\frac{\vec J}{|\vec J|},\frac{\vec x }{k}\Big)
 \,,
 &
\eeal{27V16.1}
and define $ \sigma^k_{\vec 0} : =0= : \tau^k_{\vec 0}$. We will often write $\sigma^k$ for $\sigma^k_{\vec J}$, with $\vec J$ implicitly understood, similarly for $\tau^k$.
Let $\gim$ be as in \eq{p12V16.12} and
introduce
\bel{20V16.3}
 (\hat g, \hat K)\equiv ( \gim
 +\frac 1 k \sigma_{\vec J}^k
%  +\frac 1 {4k} \hat \sigma_{\vec c}^{4k}
 ,
%  \Kivep  +
\frac 1 {k^2} \tau_{\vec J}^k)
 \,.
\ee
%%
%
%Some comments about the prefactors $k^{\lambdac-1/2}$ and $k^{1/2-\lambdac }$ in \eq{27V16.1} are in order. One can think of the $\sigma_1$-contribution in \eq{27V16.1} as adding mass to the initial data set, and the $\tau_1$-contribution as giving velocity to this mass. The prefactors involving powers of $k$ are chosen to cancel out, so that the net product mass times velocity is independent of the prefactors.  It is precisely this balance which allows us to reach the optimal weighted H\"older decay rate $\alpha_->1/2$ in the proof. Allowing a larger exponent in the  prefactor on $\sigma_1$ would lead to infinite mass; decreasing the exponent of the prefactor on $\sigma_1$ at the price of increasing the exponent of the prefactor on $\tau_1$ (e.g., not putting any prefactors at all) would lead to infinite kinetic energy unless one restricts further the range of $\alpha_-$.
%
We will only consider $m$ and $\vec J$ such that
\bea
&
  \frac \varepsilon 2 |\mathring m|\le |m| \le 2\varepsilon |\mathring m|
 \,,
 \quad
 %
%  |\vec p| \le 2 \varepsilon |\vec{\mathring p}|
%  \,,
%  \quad
% |\vec c\normwithnodelta   +
|\vec J\normwithnodelta
 \le \nolambda
 k^{2-2\alpha_c}
  \,.
  &
\eeal{12V16.3}
Let us denote by
\bel{27V16.4}
 (\R^3, \og, \ol K) \equiv
 (\R^3,\gveQk , \KvepQk )
\ee
the family of vacuum initial data of the form
\bel{27V16.2}
 \og = u^4 \hat g
 \,,
 \quad
 \ol P_{ij} = u^2
  (\hat P_{ij} + \hat D_i X_j  + \hat D_j X_i - \hat D^k X_k \hat g_{ij})
 \,,
\ee
where $\hat D$ is the covariant derivative operator of the metric $\hat g$,
with
\bel{2IX16.3}
 \ol P_{ij} = - \ol K_{ij} + \tr_\og \ol K \og_{ij}
 \,,
 \quad
 \hat P_{ij} = - \hat K_{ij} + \tr_{\hat g} \hat K \hat g_{ij}
 \,,
\ee
and
where $u$ and $X$ are obtained by solving the constraint equations for $(g,K)$ of the form \eq{27V16.2}.
These are essentially the same as  the initial data used in~\cite{HSW} (the compactly supported additions in~\cite{HSW,CorvinoSchoen2} are not needed here as there will be no cokernel in our case). However, we have to reexamine the construction of \cite{HSW} because of the need to use pairs $(\vec c,\vec J)$ which are allowed to grow in norm with $k$, cf.\ \eq{12V16.3}. Indeed, it is not even clear whether the required solutions of the constraint equations exist with the needed ranges of parameters.

Thus, we  view the  constraint operator  as a functional of $u$ and $X$:
\bea
 \mathcal C(u,X) :=
  \left(
  \begin{array}{c}
    (\Delta_{\hat g} u - \frac{\hat R}{8})u
     + \frac 1 8 ( |\ol K|_{\ol g} ^2  -  (\tr_{\ol g} \ol K)^2)u^5
      \\
   2 \overline D^i\big(u^{2}(\hat D_i X_j + \hat D_j X_i - \hat D^k X_k \hat g_{ij} + \hat P_{ij})\big)
  \end{array}
  \right)
   \,,
\eeal{29V16.4}
where $\ol D$ is the covariant derivative operator of the metric $\ol g$, and where in the first line $ \ol K$ should be expressed in terms of $X$ and $\hat P$ as in \eq{27V16.2}-\eq{2IX16.3}.

The fields \eq{20V16.3} fail to satisfy the vacuum constraints only on the annulus $A(k)$.
Since $\sigma^k$ and $\tau^k$ satisfy the linearised constraint equations in the Euclidean metric $\delta$, on $A(k)$ the violation of the scalar constraint by the data $(\hat g, \hat K)$, which will be denoted by $\hat{\mathcal C_s}$,
is a sum of terms such as $k^{-1}\partial \sigma^k \partial   \gim  $ , $k^{-2}(\partial \sigma^k)^2 $, %, $k^{-2}\Kivep   \tau^k$
 $ k^{-4}( \tau^k)^2$, and somewhat similar, which can all be estimated as%
%%
%\bean
% \lefteqn{
% \|{\hat{\mathcal{C}_s}}\|_{W^{-\beta-2}_{\ell-2 ,q}}^q
% \le
%    \int_{\fromktotwok } (1+r)^{-n}
%    \times
%    }
%    &&
%\\
% \nn
%  &&
%      \Big((1+r)^{\beta+2}
%   \big( C \nolambda k^{\frac 12 -\lambdac}
% + k^{-\frac 12 + \lambdac} |\vec J|
% +  k^{-1 + 2 \lambdac}|\vec J|^2\big)O(k^{-4} )\Big)^q
%  d^3 x
%\\
%  & \le &
%   C
%   \Big(\nolambda k^{\beta+2} \big(     k^{-\frac 72 -\lambdac}
% + k^{ -\frac5 2 - 2 \alpha_c + \lambdac}
% +  k^{-1-4\alpha_c + 2 \lambdac} \big)\Big)^q
%  \nn
%\\
% & \le &
%   3 C \nolambda k^{(\beta -1 - \deltac)q}
%% \,,
%\eeal{20V16.5}
%%
%
\bean
 \lefteqn{
 \|{\hat{\mathcal{C}_s}}\|_{W^{-\beta-2}_{\ell-2 ,q}}^q
% &
 \le
% &
  C   \int_{\fromktotwok } (1+r)^{-3}
   \times
    }
    &&
\\
 \nn
  &&
 \Big(
  (1+r)^{\beta+2} k^{-4 }
   \big(
% + k^{-\frac 12 + \lambdac}
 |\vec J|^{\frac 12} +
%k^{-4}
|\vec J|
 \big)
  \Big)^q
  d^3 x
%\\
%  & \le &
%   C
%   \Big(\nolambda k^{\beta+2} \big(     k^{-\frac 72 -\lambdac}
% + k^{ -\frac5 2 - 2 \alpha_c + \lambdac}
% +  k^{-1-4\alpha_c + 2 \lambdac} \big)\Big)^q
%  \nn
%\\
% & \le &
 \le
    C \nolambda k^{(\beta -2 \alpha_c)q}
     =: C k^{-q\deltac}
      \,.
%% \,,
\eeal{20V16.5}
%
%where, keeping in mind that $\lambdac$ can be chosen as small as desired,
%%
%\bel{24VIII16.1}
% 0<\deltac:=\min(7/2+\lambdac, 5/2 +2 \alpha_c-\lambdac,1+4\alpha_c-2\lambda_c)-3<1
%  \,.
%\ee
%%
For any $0<\beta<1$ %we can choose $q$ large enough so that $\deltac>0$, therefore
the norm in \eq{20V16.5} goes to zero as $k$ tends to infinity.

An estimate  for the violation of the vector constraint, say $
\hat {\mathcal C}_v$, can be similarly derived:
\bea
&
 \|{\hat{\mathcal{C}_v}}\|_{W^{-\beta-2}_{\ell-2 ,q}}
  \le
     C \nolambda k^{ \beta -2\alpha_c }
  =
     C \nolambda k^{ - \deltac }
 \,.
 &
\eeal{29VIII16.1}
%
%after reducing $\deltac$ if necessary.

Let $L$ denote the linearisation of $\mathcal C$ at the initial data $(\hat g,\hat K)$, $u\equiv 1$, and $X\equiv 0$. Standard considerations (cf., e.g.,~\cite{ChoquetBruhatChristodoulou81}) show that  for all $\varepsilon$ sufficiently small all the operators
\bel{2IX16.1}
 L: W^{-\beta}_{\ell,q}\mapsto W^{-\beta-2}_{\ell-2,q}
 \,,
 \quad
 \ell \ge 2\,,
 \ \beta \in (0,1)
 \,,
 \
 q\ell> 3
 \,,
%  \
%   \deltac >0
%  \,,
\ee
obtained by varying $m $ and $\vec J$ as in \eq{12V16.3} and $k\in \N$ are  isomorphisms  for
%each $\beta\in (0,1)$ and for
all $k$
%and $q$
large enough, with the  norms of their inverses  bounded independently  of $k$, $\varepsilon$, $ m$ and $\vec J$ within the ranges considered.
It follows that we can apply the implicit function theorem to the equation ${\mathcal C}(u,X)=0$ to obtain existence of  solutions%
\footnote{More precisely, for any parameters as in \eq{2IX16.1} there exists a solution for $k$ large enough. Uniqueness implies that the solutions are independent of the triples $(\beta,q,\ell)$. So the  solution found for, say, $\beta=1/2$, $q=\ell=100$  satisfies the estimates claimed for all $\beta\in(0,1)$, $q>3/\ell$ and $\ell\ge 2$.}
$$
 (u,X)\in W^{-\beta }_{\ell ,q}\,,
 \quad
  \beta\in(0,1)
  \,,
$$
satisfying
\bel{20V16.5+2}
 \| (u,X)\|_{W^{-\beta }_{\ell  ,q}}
 \le
   C
    \nolambda k^{-\deltac}
   \,.
\ee
Here one should keep in mind that such a weighted Sobolev estimate implies weighted pointwise decay estimates for the solution and its derivatives, e.g.
\bel{20V16.5+3}
 |u|+|X_i| \le\frac{ C \| (u,X)\|_{W^{-\beta }_{\ell  ,q}} }{(1+r)^\beta}
 \le
 \frac{ C^2
    \nolambda k^{-\deltac}}{(1+r)^\beta}
   \,.
\ee

We will use the symbol  $\hepsilon$ to denote a small positive constant which can vary from line to line. In the calculations that follow it is convenient to note that we can write
$$
 k^{-\beta-\deltac} \equiv k ^{-2\alpha_c} = O( k^{-1-\hepsilon})
 \,,
$$
%
%with $\hepsilon>0$,
%after choosing $\beta$ close enough to one,
and that $r^{-\beta} = O(r^{-1+\hepsilon})$, $r^{-2\beta} = O(r^{-2+\hepsilon})$, etc., making the constant $\hepsilon$ smaller at each further equality if necessary.

Let $L_s$ denote the linearisation with respect to $u$, at $u\equiv 1$, of the operator appearing in the upper line of \eq{29V16.4}. Set
$$
 \hat u:= u -1
 \,.
$$
%.
We will have $1/2\le u \le 2$ for $k$ large enough, which implies that the function $\hat u$ satisfies an equation of the form
\beal{31VIII16.1}
 L_s \hat u
  & = &
    O (|\hat P   |_{\hat g}^2)
  )
\\
  & = &
    O\Big( \frac{k^{-\hepsilon}}{(1+r)^{4-\hepsilon}} + k^{-3-\hepsilon} \one_{A(k)}
  \Big)
  \,,
\eeal{31VIII16.8}
with
\beal{31VIII16.2}
 L_s
  & = &
   \Delta_{\hat g}  - \frac {\hat R} 8  + O(|\hat P|_{\hat g}^2)
\\
  & = &
   \Delta_{\hat g}
   + O
     \Big(
     \frac{k^{-\hepsilon}}{(1+r)^{4-\hepsilon}} + k^{-3-\hepsilon} \one_{A(k)}
  \Big)
  \,,
\eeal{31VIII16.3}
A standard asymptotic analysis, using  e.g.~\cite{ChAFT}, shows that for $r>1$ we have
\bel{31VIII16.5}
 \hat u = \frac b r + \hat u_1
  =: \frac b{\sqrt{1+r^2}} + \hat u_2
 \,,
\ee
where $b$ is a constant,
with
\bel{31VIII16.5+}
  b  = O(k^{-\hepsilon})
  \,,
  \quad
    \hat u_2 = O\Big( \frac{k^{-\hepsilon}}{(1+r)^{2-\hepsilon}}
    \Big)
 \,,
\ee
and with the last estimate holding on $\R^3$.
Inserting this into the equation satisfied by $X$ we find
\bean
\lefteqn{
 \hat D^i\big( \hat D_i X_j + \hat D_j X_i - \hat D^k X_k \hat g_{ij} \big)
% =
}
 &&
  \nn
\\
 &  &
 =
  -2 u^{-2}\hat D^i\hat P_{ij}
  + O(\hat D  u \times  \hat  D X )
%  \label{31VIII16.9}
% & = &
=
    O\Big( k^{-2-2 \alpha_c} \one_{A(k)} + \frac{k^{ -\hepsilon}}{(1+r)^{4-\hepsilon}}
  \Big)
 \nn
\\
 &  &
 =
    O\Big( k^{-3 -\hepsilon} \one_{A(k)} + \frac{k^{ -\hepsilon}}{(1+r)^{4-\hepsilon}}
  \Big)
  \,.
\eeal{31VIII16.10}
where we have again used the fact  that $ \tau_{\vec J}^k $ satisfies the linearised vector constraint in the Euclidean metric.
This implies, for $r\ge 1$,
\bel{31VIII16.11}
  X_i= \frac {c_i} r    + \hat X_i=: \frac{c_i}{\sqrt{1+r^2}} + \tilde X_i\,,
  \qquad
   |\tilde X_i| =  O\Big(  \frac{k^{ -\hepsilon}}{(1+r)^{1+\hepsilon}}
   \Big)
   \,.
\ee
with some constants $c_i$.

From the fact that both $\hat g$ and $\hat K$ are even under the parity map, using uniqueness of solutions we infer that
$u$ is even and the $X_i$'s are odd (note that the one-form $X_idx^i$ is thus  parity-even under pull-back), hence
\bel{29VIII16.6}
 c_i =0
 \quad
 \Longrightarrow
 \quad
  | X_i| =  O\Big(  \frac{k^{ -\hepsilon}}{(1+r)^{1+\hepsilon}}
   \Big)
 \,.
\ee

For $|\vec x| \ge \reallyfourk $ the equation  $\mathcal C(u,X) =0$ reads
\bea
0 =
  \left(
  \begin{array}{c}
    \Delta_{\hat g} u
     + \frac 1 8 ( |\ol K|_{\ol g} ^2  -  (\tr_{\ol g} \ol K)^2)u^5
     \\
   2 \overline D^i\big(u^{2}(\hat D_i X_j + \hat D_j X_i - \hat D^k X_k \hat g_{ij} )\big)
  \end{array}
  \right)
   \,.
\eeal{29V16.4n}
One can insert in \eq{29V16.4n} the improved estimates \eq{31VIII16.5+} and \eq{29VIII16.6} to obtain,
 again by standard arguments and taking the parity properties of $(u,X)$ into account, that $(u,X)$ have full asymptotic expansions in terms of powers of $r^{-1}$ and $\ln r$, with
\bean
 &
 \displaystyle
 u = 1 + \frac b r   + \tilde u\,,
 \quad
 X_i=  \frac { c_{ij}   x^j} {r^3}  + \tilde X\,,
 &
\\
 &
 \tilde u \,,
 \hat X_i \in C^{-2-\epsilonc}_{\ell+2 }(\R^3\setminus B(\twok ))
 \,,
 &
\eeal{p27V16.9}
where the   $c_{ij}$'s are constants, with%
\bel{29VIII16.8}
  | \tilde u | + |  \tilde X_i | \le C \frac{k^{ -\epsilonc }} {r^{2+\epsilonc}}
 \,,
\ee
in fact

\bel{29VIII16.9}
 \| \tilde u \|_{ C^{-2-\epsilonc}_{\ell+2 }}
  + \|
 \tilde X_i\|_{ C^{-2-\epsilonc}_{\ell+2 }}\le C k^{ -\epsilonc }
 \,.
\ee

The parity properties of $u$ and $X$ imply that both $\ol g$ and $\ol K$ are even. In particular the center-of-mass integrals  of $\ol g$ on any centered sphere vanish, similarly for the ADM momentum integrals.

Since  $b=O(k^{-\hepsilon})$, the mass integrals at $R\ge \dwa k$ of $(\ol g,\ol K)$ approach those of $(\hat g,\hat K)$ as $k$ goes to infinity.

It further follows from the calculations in~\cite{HSW} and the estimates above that for $R\ge 4k$ the angular momentum integrals at $R \ge \twok  $ of $(\ol g, \ol K)$ approach  the vector $\vec J$ as $k$ tends to infinity. One can then determine the constants $c_{ij}$ algebraically through $\vec J$, which leads to
\bel{31.8.16.21}
 c_{ij} = O( \nolambda k^{2-2\alpha_c})
 \,.
\ee

We are ready to finish the proof for \emph{parity-even} initial data $(g,K)$, by  gluing together  across an annulus $A(R) $, with $R=\dwa k$ and for $k$ large,  the initial data sets  $(E(R),g^{ \varepsilon }, K^{ \varepsilon })$ and $(B(\dwa R),\gveQk , \KvepQk )$, as in~\cite{CCI2}. The boundary terms at $\{|\vec x|=1\}$ and  $\{|\vec x|=\dwa \}$ in the rescaled balance formula \eq{p12VIII16.3} for mass and angular momentum can be calculated by scaling the boundary terms of the metrics $(B(2 R),\gveQk , \KvepQk )$ and $(E(R),g^{ \varepsilon }, K^{ \varepsilon })$. The error terms are rather similar to those that occur in the calculations already carried-out in the time-symmetric case, we leave the details to the reader. The fact that the Sobolev norms of the resulting initial data sets go  to zero as $\varepsilon$ goes to zero  follows directly from the estimates established so far.

For general $(g,K)$ we need to enlarge the family of interior candidates to obtain initial data sets with center of mass and angular momentum in the relevant ranges. For this, let
\bel{3IX16.10}
 (E(R),\gKKerr)
\ee
be initial data on a slice $x^0=0$  in Kerr-Schild coordinates, as in~\cite[Section~2.1]{CCI2}.  Since the data \eq{3IX16.10} are parity-symmetric, we can carry out the gluing just described of  $(E(R),\gKKerr)$, with $R=\dwa k$, together with $(B(\dwa R= \cztery k),\gveQk  , \KvepQk  )$ on $A(\dwa k)$. The resulting initial data on $\R^3$ have small energy and therefore exist globally in harmonic coordinates by~\cite{LindbladRodnianski2}. The solution is exactly Kerr in the domain of dependence of $E(\cztery k)$.
%We denote the resulting initial data by
%%
%$$
%\idmJk
% \,.
%$$

After translating by a vector $- \vec a$, satisfying $ |\vec a| \le   k^{2-2\alpha_c}<k$,
one obtains initial data which are exactly Kerrian outside of $B(\osiem k)$, have  center of mass integrals at $R\ge \osiem k$ which approach $\vec c =m \vec a$ as $k$ tends to infinity, and  angular momentum integrals at $R\ge \osiem k$ approaching
$
 \vec J
$
as $k$ tends to infinity. This provides a family of initial data
$$
 \idmJkc
$$
with small energies. Performing a Lorentz transformation on the global harmonic coordinates so that the level sets of the new time coordinates have momentum
$$
 |\vec p | %\le 4 \varepsilon |\vec{\mathring p}|
 \le 4 \varepsilon \mathring m
$$
we obtain a family of initial data
$$
 \idQk
$$
with small energies. Keeping in mind that $(\vec a, \vec J)$ transform linearly under Lorentz transformations, the family contains all $Q= (m,\vec p, \vec c,\vec J)$ with
\bel{28VIII16.4}
 |\vec a|  +
 |\vec J| \le (1+C \varepsilon \mathring m ) k^{1-2\alpha_c}
 \,,
\ee
for some constant $C$.

A final gluing of $\idBQk$ with $(E(\osiem k),g^\varepsilon,K^\varepsilon)$ across $A(\osiem k)$ provides the desired vacuum initial data.
This completes the proof of Proposition~\ref{P27VIII16.2}.
\qed

\bigskip

\noindent{\sc Acknowledgements:}  I am  grateful to the Center for Mathematical Sciences and Applications at Harvard University for hospitality and support during part of work on this paper. Supported in part by the Austrian Science Fund (FWF)  project  P29517-N16.

\bibliographystyle{amsplain} 
%\end{document}
\bibliography{%
../references/reffile,%
../references/newbiblio,%
../references/hip_bib,%
../references/newbiblio2,%
../references/bibl,%
../references/howard,%
../references/bartnik,%
../references/myGR,%
../references/newbib,%
../references/Energy,%
../references/netbiblio,%
../references/PDE}

\end{document}